
\input phyzzx
\pubnum{ROM2F-95-29}
\def\a{\alpha}
\def\b{\beta}

\def\d{\delta}

\def\m{\mu}
\def\n{\nu}
\def\r{\rho}
\def\s{\sigma}
\def\t{\tau}

\def\mnr{\mu\nu\rho}
\def\mn{\mu\nu}

\def\wh{\widehat}
\def\wt{\widetilde}
\def\de{\partial}

\def\mezzo{{1 \over 2}}

\def\ap{\alpha^\prime}
\def\unita{{1 \kern-.30em 1}}
\def\fey{{\big / \kern-.80em D}}
\def\complex{{\kern .1em {\raise .47ex \hbox {$\scriptscriptstyle |$}}
\kern -.4em {\rm C}}}
\def\zet{{Z \kern-.45em Z}}
\def\real{{\vrule height 1.6ex width 0.05em depth 0ex
\kern -0.06em {\rm R}}}
\def\rational{{\kern .1em {\raise .47ex \hbox{$\scripscriptstyle |$}}
\kern -.35em {\rm Q}}}

\titlepage
\title{Duality Symmetries in String Theory \foot{
Expanded version of the talk presented
at the IV Italian-Korean meeting on Relativistic Astrophysics,
Rome - Gran Sasso - Pescara, July 9-15, 1995}}
\author{Massimo BIANCHI}
\address{Dipartimento di Fisica, Universit\`a di Roma II ``Tor
Vergata''
I.N.F.N. Sezione di Roma II ``Tor Vergata'', 00133 Roma, Italy}

\abstract
{We review the status of duality symmetries in superstring theories.
These discrete symmetries mark the striking differences between theories
of pointlike objects and theories of extended objects.
They prove to be very helpful in understanding
non-perturbative effects in string theories.
We will also briefly discuss the strange role played by open strings and their
solitons in the emerging scenario.}
\endpage
\pagenumber=1

\chapter {Introduction}

\REF\gpr{A.~Giveon, M.~Porrati and E.~Rabinovici,
Phys.Rep. {\bf 244} (1995) 77.}
In the mounting wave of interest in string theory a key role is played
by discrete symmetries (dualities) [\gpr].
This talk is intended to be an introduction for non experts to these issues.
In the discussion we will try to show how these dualities
may prove useful in the study of non-perturbative effects in string theory.
We will also emphasize the
dramatic consequences implied by the presence of duality symmetries
in the interpretation of
low-energy effective field theories which are stringy generalizations
of General Relativity.

\REF\thv{G.~'t~Hooft and M.~Veltman, Ann. Inst. H.~Poincar\`e 20 (1974) 69.}
\REF\gs{M.~Goroff and A.~Sagnotti, Nucl.Phys. {\bf B266} (1986) 709.}
Classical General Relativity is a very successful theory for the description of
gravity at large distances. However it is well known that at very short
distances, \ie~at energy comparable with the Planck mass $M_P=(hc/G_N)^{1/2}$,
with $G_N$ the Newton constant,
the theory is inconsistent with quantum mechanics. Technically speaking a
quantum theory of gravity based on the Einstein-Hilbert action
$$
S = {1\over 16\pi G_N} \int R(g) \sqrt{g} d^4x
\eqn\ehaction
$$
is not renormalizable since the coupling constant of the theory, $G_N$,
has dimension of an inverse mass squared.
In natural units indeed $G_N=M_P^{-2}$, and though there are no
on-shell divergences in pure gravity at one-loop [\thv], at two-loops an
on-shell counterterm is generated which signals the quantum inconsistency
of the theory [\gs].
\REF\ffvn{S.~Ferrara, D.~Freedmann and P.~Van Nieuwenhuizen, Phys. Rev.
{\bf D13} (1976) 3214.}
\REF\ggrs{S.~Gates, M.~Grisaru, M.~Rocek and W.~Siegel, {\it Superspace},
Benjamin Cummings Readings, Mass. 1983.}
In almost any kind of coupling of gravity to matter fields things worsen
(the counterterm which can be generated at one-loop does not vanish
on-shell any more), except when the coupling is supersymmetric.
The ``miraculous" cancellations between boson and fermion loops, which make
(rigid) supersymmetry the only viable candidate for the
resolution of the hierarchy problem, persist at low orders in locally
supersymmetric theories, known as (extended) supergravity [\ffvn].
The resulting theories are still non-renormalizable and one expects
divergencies
to be generated at three loops or more [\ggrs].

\REF\gsw{M.~Green, J.~Schwarz and E.~Witten,
{\it Superstring theory}, Cambridge University Press 1987.}
In order to describe gravity at short distances one needs a more radical
change of perspective. Thanks to its very soft ultraviolet
behaviour (``Regge behaviour")
string theory is the only known candidate to the unification
of gravity with the other interactions in a consistent quantum theory
(for a review see [\gsw]).
There are two classes of string theories: those with only closed
oriented strings and those with open and closed unoriented strings.
The string spectrum is encoded in the Regge trajectories which relate the
masses $M$ to the spins $J$ of the states. For the closed-string states
$$
\ap M_J^2 = J - 2
\eqn\closedspec
$$
which implies the existence of a massless tensor field (``graviton").
For the open-string states the relation becomes
$$
\ap M_J^2 = J - 1
\eqn\openspec
$$
and one may notice the presence of a massless vector boson (``photon").
In the early days of the hadronic string,
in order to match the string states with the known resonances,
the string tension $T=1/2\pi\ap$
was supposed to be of the order of the proton mass $T=M_p^2$.
In the unified string picture the string tension is taken to be comparable
with the Planck mass squared $T=M_P^2$ and the tower of massive states
decouples from the massless states at low energies,
\ie~in the limit $\ap\rightarrow 0$.
The infinite tower of massive modes, which at low energies can be neglected or
better ``integrated out'', may play a significant role in the
early stages of evolution of the universe. The exponential growth
of the degeneracy of string states with energy signals a phase
transition at the Hagedorn temperature ($T_H \approx 1/\sqrt{\ap}$).

In order to quantize string theory and to display its symmetry principles,
it is very convenient to proceed in analogy with the Feynman path-integral
quantization of a point-particle. Instead of summing over ``world-lines'',
Polyakov proposed to sum over ``world-sheets", \ie~surfaces spanned by the
string in its time evolution. The proper weight is the action of Brink, Di
Vecchia and Howe (BDVH)
$$
S[X,\gamma] = {1\over 4\pi\ap}
\int_\Sigma G_{\mn}(X) \de_\a X^\m \de_\b X^\n \sqrt{\gamma}
\gamma^{\a\b}d^2\sigma
\eqn\polaction
$$
where the string coordinates $X^\m$ are maps from the world-sheet $\Sigma$
to a ``target space" ${\cal M}$ with metric $G_{\mn}$.
In order to compute amplitudes and get meaningful results one has to divide
out the volume of the huge symmetry group of the classical action.
The BDVH action is invariant under two-dimensional diffeomorphisms as well as
under Weyl rescalings of the world-sheet metric $\gamma_{\a\b}$.
Using the freedom of world-sheet coordinate changes
one can fix $\gamma_{\a\b}$ to
be conformally flat $\gamma_{\a\b} = e^{2\s} \d_{\a\b}$.
At the classical level, the conformal factor $\s$ decouples.
At the quantum level, conformal anomalies are generated upon
removing an explicit regularization scale. Following
the standard Faddeev-Popov (FP) procedure one finds that the conformal anomaly
of the string coordinates $X^{\m}$ is cancelled
by the contribution of the FP ghosts when $D=26$, which is to be considered
as the critical dimension for the bosonic string. An analogous
procedure shows that the critical dimension for
the superstring, which requires the introduction of world-sheet
fermionic partners
$\Psi^\m$ of the bosonic coordinates $X^{\m}$ as well as the superconformal
FP ghosts, is $D=10$.
Using conformal invariance the Polyakov integral may be reduced to
a perturbative series in the topology of Riemann surfaces. Once the
functional integration over the fields $X^\m$ and $\Psi^\m$ has been performed,
one is left with a finite dimensional integral over the Teichm\"uller
parameters
which describe the ``shape" of a Riemann surface.
The measure of integration descends to the moduli space of Riemann surfaces
if invariance under
``large" (disconnected) diffeomorphisms, generating the mapping class group
(often called the modular group), is preserved at the quantum level.
To summarize, conformal invariance and modular invariance are the guiding
principles for the construction of (perturbatively) consistent string models.

\chapter {Superstrings in $D=10$}

The known supersymmetric and, as such, tachyonic free string theories can all
be naturally formulated in $D=10$.
According to the number of (Majorana-Weyl) supersymmetries one has:

{\it Type I superstring} - a theory of open and closed unoriented
superstrings whose low-energy limit is the chiral $N=(1,0)$ supergravity
coupled
to $N=(1,0)$ super Yang-Mills theory with gauge group $SO(32)$;

{\it Type IIA superstring} - a theory of closed
oriented superstrings, L-R asymmetric on the world-sheet, whose low-energy
limit
is the non-chiral $N=(1,1)$ supergravity;

{\it Type IIB superstring} - a theory of closed
oriented superstrings, L-R symmetric on the world-sheet, whose low-energy
limit is the chiral $N=(2,0)$ supergravity;

{\it Heterotic string} - a L-R asymmetric combination of bosonic and
fermionic strings whose low-energy limit is the chiral $N=(1,0)$ supergravity
coupled to $N=(1,0)$ super Yang-Mills theory with gauge group $Spin(32)/Z_2$
or $E(8)\times E(8)$.

The first and the last of these theories, though chiral, are anomaly free
thanks to
the Green-Schwarz mechanism which involves the antisymmetric tensor field
$B_{\mn}$ present in the massless spectrum. The type IIB theory, though
chiral, is anomaly free thanks to a remarkable cancellation among the
contributions of two left-handed gravitini, two right-handed
dilatini and a four-form potential $A^{(+)}_{\m\n\r\s}$ whose field
strength is to be taken self-dual.
All the above theories contain a metric $G_{\mn}$,
an antisymmetric tensor $B_{\mn}$ and a scalar (dilaton) $\Phi$
as massless particles in their bosonic spectra.
In addition, in the open-string spectrum of the type I theory and in the
Neveu-Schwarz (NS) spectrum of the
heterotic theory massless vector bosons $A^a_\m$ appear.
In the type II theories the massless NS-NS
spectrum is accompanied by a rich spectrum of tensor fields
coming from the massless Ramond-Ramond (R-R) sector.
The type IIA contains an abelian vector field $A_\m$ and a three-form
potential $C_{\mnr}$, while the type IIB theory, apart from the
above discussed $A^{(+)}_{\m\n\r\s}$, contains
another antisymmetric tensor $B'_{\mn}$ and a second dilaton $\Phi'$.
In the Polyakov approach,
the vacuum expectation value of the dilaton field plays the role of string
coupling constant, \ie~of string-loop expansion parameter.

\chapter{Superstring Compactifications and T-duality}

The superstring theories formulated in $D=10$ are not very appealing
as models for particle physics. One would like to find a mechanism
to compactify the unwanted six dimensions in order to remain with only
four physical dimensions. Unfortunately string theory admits so far only
a perturbative formulation and
a dynamical principle to infer compactification is still lacking.
All one can do is to study the condition for perturbatively consistent
compactifications
and to deduce the spectrum and interactions of the resulting lower dimensional
theories.
The stringy generalization of the dimensional reduction \'a la Kaluza-Klein
reveals some unexpected features. Indeed apart from the usual Kaluza-Klein
(K-K)
modes with quantized momenta, there exist ``winding"
modes corresponding to the string wrapping around the compact dimensions.
The presence of these extra modes is responsible for the so-called
Halpern-Frenkel-Kac (H-F-K) mechanism of symmetry enhancement.
It is well known that in K-K theories the components of the metric
with mixed indices $G_{\m i}$ behave as vector bosons
which gauge the isometry group of the internal manifold.
For instance, the K-K gauge symmetry arising from $k$ abelian
isometries of a $k$-torus
$T^k = S^1\times S^1 \dots S^1$ can be enlarged in string theory to
a non-abelian group, for special
values of the compactification radii. More specifically, the spectrum of the
string states
corresponding to a one-dimensional compactification on a circle of radius $R$
is given by
$$
M^2_\pm = M^2_J + \mezzo ({m\over R} \pm {nR\over \ap})^2
\eqn\winding
$$
were $\pm$ denote L(R)-moving modes, $m/R$ is the quantized momentum and
$n$ is the winding number. The spectrum shows a striking
symmetry under the exchange $R\rightarrow \ap /R$ and at the fixed
point of the transformation, \ie~$R=\sqrt{\ap}$, new massless fields appear
in the spectrum (those with $m=\pm n =\pm 1$) that enlarge the $U(1)$
symmetry to $SU(2)$. In fact in closed oriented string theories there
is a doubling of modes and both the abelian and enhanced symmetries get
doubled.
In the low-energy description the vector
fields which arise from the mixed components of the antisymmetric tensor
$B_{\m i}$ are responsible for this doubling.
In more complicated situation the above symmetry between large and small
radius becomes an infinite dimensional discrete non-abelian symmetry
which acts on the vacuum
expectation values (v.e.v.'s) of the internal components of the metric
$G_{ij}$ and of the
antisymmetric tensor $B_{ij}$ in such a way as to preserve the string spectrum
\winding. From the low-energy point of view, $G_{ij}$ and $B_{ij}$ correspond
to
massless scalars with only derivative interactions.
Their v.e.v.'s parametrize the space of classical vacua usually called
the moduli space of the compactification. These moduli spaces are
(locally) coset manifolds which admit very large group of isometries.
Due to non-perturbative world-sheet effects
(such as the windings) the global continuous symmmetries
are reduced to discrete groups of so-called ``T-duality".

The massless spectrum of type II theories compactified
on a flat six-dimensional torus reproduces $N=8$ supergravity.
It includes 28 abelian vector bosons, called graviphotons (12 from the
NS-NS sector and 16 from the R-R sector), and
70 scalars which parametrize the coset $E(7,7)/SU(8)$.
The T-duality group is $SO(6,6;Z)$. For heterotic
theories one has $N=4$ supergravities coupled to non-abelian vector multiplets.
However at generic points of the moduli space, $SO(6,22)/SO(6)\times SO(22)$,
the non-abelian group is broken to $U(1)^{28}$ and the T-duality group is
$SO(6,22;Z)$. For the type I superstring, which leads to $N=4$ low energy
effective field theories, one would expect a behaviour similar to the
heterotic string case, but in the perturbative string spectrum part of the
T-duality symmetry is absent.
The asymmetric origin of the scalar fields, which are
the internal components of the metric $G_{ij}$ (from the closed string NS-NS
sector), the internal component of the antisymmetric tensor $B_{ij}$ (from the
closed string R-R sector) and the internal components of the gauge fields
$A^a_i$ (from the open string sector) prevents
the existence of any obvious transformation among them.
Supergravity considerations lead one to suspect that the inclusion of solitonic
states, charged with respect to the R-R vector bosons, may restore the
full T-duality symmetry.

\chapter{Buscher's duality and Stringy Gravitational Instantons}

\REF\bus{T.H.~Buscher, Phys. Lett. {\bf B194} (1987) 51; ibid. {\bf 201}
(1988) 466.}
The T-duality symmetry between large and small radius in toroidal
compactifications has an analogue in any background with continuous isometries.
Indeed Buscher [\bus] has shown that if a two dimensional non-linear
$\s$-model,
describing string propagation in a non trivial background
with metric $G_{ij}(X)$, antisymmetric tensor $B_{ij}(X)$ and dilaton
$\Phi(X)$, admits an abelian isometry then it is equivalent to
a dual $\s$-model on a background
with metric $\tilde G_{ij}$, antisymmetric tensor
$\tilde B_{ij}$ and dilaton
$\tilde\Phi(X)$. In adapted coordinates, such that the Killing vector is
$\xi=\de/\de X^o$, the relation between the two backgrounds reads
$$
\eqalign{
\tilde G_{oo} &= {1\over G_{oo}} \quad
\tilde G_{oi} = {B_{oi}\over G_{oo}} \quad
\tilde G_{ij} = G_{ij} - {1\over G_{oo}} (G_{oi}G_{oj} - B_{oi}B_{oj})\cr
\tilde B_{oi} &= {G_{oi}\over G_{oo}} \quad
\tilde B_{ij} = B_{ij} - {1\over G_{oo}} (G_{oi}B_{oj} - B_{oi}G_{oj}) \quad
\tilde \Phi = \Phi - \mezzo logG_{oo}\cr}
\eqn\buscher
$$
\REF\aagbl{E.~Alvarez, L.~Alvarez-Gaum\'e, J.~Barbon and Y.~Lozano,
Nucl. Phys. {\bf B415} (1994) 71.}
The above transformations have dramatic consequences for the geometric and even
for the topological interpretion of string backgrounds [\aagbl].
For example, in black-hole type
geometries, Buscher's duality \buscher~exchanges singularity and horizon.

\REF\dkl{M.~Duff, R.~Khuri, J.~Lu, Phys.Rep. {\bf 259} (1995) 213.}
The very consistency
of string propagation on a non-trivial background imposes the vanishing of the
$\b$-functions of the corresponding $\s$-model. They play the role of string
equations of motion and are equivalent to extremizing an action functional
for the effective low-energy field theory.
Non trivial supersymmetric solutions of the lowest order (in $\ap$) equations
of motion may be found by setting to zero the fermion fields together with
their
supersymmetric variations (for a review see [\dkl]).
In $D=4$ a supersymmetric ansatz for the solution of the heterotic string
equations of motion is
$$
F_{\mu\nu} = \wt F_{\mn} \qquad
H_{\mnr}= \sqrt{G} \varepsilon_{\mnr}{}^\sigma\de_\sigma\phi \qquad
G_{\mn} =  e^{2\phi} \wh g_{\mn} \qquad
\eqn\ansatz
$$
where $F_{\mn}$ is the field-strength of the gauge fields $A_\m$, $H_{\mnr}$
is the (modified) field-strength of the antisymmetric tensor $B_{\mn}$
and $\wh g_{\mn}$ is a self-dual metric. The generalized spin-connection
with torsion $\Omega_{\m\pm}^{ab}=\omega_\mu^{ab}\pm H^{ab}_\mu$
deriving from \ansatz~ is self-dual. The ansatz in
\ansatz~ must be supplemented with the (modified) Bianchi identity:
$dH=\ap \{ trR(\Omega_{-})\wedge R(\Omega_{-})-trF(A)\wedge F(A) \}$.
\REF\rey{S.-J.~Rey, Phys.Rev. {\bf D43} (1991) 526.}
\REF\chs{C.~Callan, J.~Harvey and A.~Strominger,
Nucl.Phys {\bf B359}(1991) 611; ibid. {\bf B367} (1991) 60.}
\REF\bfmrale{M.~Bianchi, F.~Fucito, M.~Martellini and G.C.~Rossi,
Nucl.Phys. {\bf B440} (1995) 129.}
To simplify matters it is convenient to impose the ``standard embedding''
of the gauge connection in the $SU(2)$ spin group: $A=\Omega_-$.
Then there are two options. The first corresponds to conformally flat
``axionic instantons" with $\wh g_{\mn} = \delta _{\mn}$ [\rey, \chs].
The second requires a constant dilaton and vanishing torsion and
it is completely specified by the choice of a self-dual
metric $\wh g_{\mn}$, \ie~a gravitational instanton [\bfmrale].

\REF\egh{T.~Eguchi, P.B.~Gilkey and A.J.~Hanson, Phys.Rep. {\bf 66} (1980)
213.}
\REF\hk{N.~Hitchin, Math. Proc. Camb. Phil. Soc. {\bf 85} (1979) 465;
P.~Kronheimer, J. Diff. Geom. {\bf 29} (1989) 665.}
An interesting class of self-dual metrics is given by the
Gibbons-\break Hawking multi-centre (GHMC) ansatz (see, \eg~[\egh])
$$
ds^2 = V^{-1}({\vec x}) (d\t + {\vec\omega}\cdot d {\vec x})^2 +
V({\vec x})d{\vec x}\cdot d {\vec x}
\eqn\multicenter
$$
with ${\vec \nabla}V={\vec \nabla}\times{\vec\omega}$ and
$V({\vec x}) = \epsilon + 2m \sum_{i=1}^{ k+1} {1\over\mid {\vec x}-
{\vec x}_i\mid}$.
The choice $\epsilon=0, m=\mezzo$ corresponds to metrics
which are asymptotically locally euclidean (ALE).
ALE instantons are smooth resolutions of singular varieties
in $\complex^3$ which admit (non-compact) Ricci-flat hyperk\"ahler metrics of
$SU(2)/\Gamma$ holonomy, $\Gamma$ being any kleinian subgroup of $SU(2)$ [\hk].

\REF\dsb{S.W.~Hawking and C.N.~Pope, Nucl.Phys. {\bf B146} (1978) 381;
E.~Witten, Nucl.Phys. {\bf B185} (1981) 513;
K.~Konishi, N.~Magnoli and H.~Panagopoulos,
Nucl.Phys. {\bf B309} (1988) 201; ibid. {\bf B323} (1989) 441.}
\REF\bfmrtop{M.~Bianchi, F.~Fucito, M.~Martellini and G.C.~Rossi,
{\it Instanton effects in supersymmetric Yang-Mills theories on ALE
gravitational backgrounds.}, ROM2F-95-12; to appear in Phys. Lett. B.}
\REF\kn{P.~Kronheimer and H.~Nakajima, Math. Ann. {\bf 288} (1990) 263.}
\REF\bfmradhm{M.~Bianchi, F.~Fucito, M.~Martellini and G.C.~Rossi,
{\it On the ADHM construction on ALE gravitational backgrounds},
ROM2F-95-16; to appear in Phys. Lett. B.}
ALE Instantons have played a fundamental role in Euclidean Quantum
(Super)-Gravity since instanton-dominated correlators give rise to
gravitino condensation which may trigger the dynamical breaking of local
supersymmetry [\dsb]. In the context of low energy effective
field theories for the heterotic or type I superstrings, the combined effects
of ALE and Yang-Mills instantons may be studied [\bfmrtop]
thanks to the possibility of generalizing the ADHM construction
[\kn, \bfmradhm].

Since \multicenter~admits a Killing vector $\xi=\de /\de \tau$
one can perform a Buscher's duality to a new background with
$$
\eqalign{
\tilde G_{oo} &= V \quad
\tilde G_{oi} = 0 \quad
\tilde G_{ij} = V \delta_{ij} \cr
\tilde B_{oi} &= \omega_i \quad
\tilde B_{ij} = 0 \quad \tilde \Phi = \Phi_o + \mezzo logV\cr}
\eqn\aledual
$$
The new background is singular and conformally flat\foot{The canonical
Einstein metric
is related to the string $\s$-model metric through a Weyl rescaling
$g_{\mn}=e^{-{4\Phi \over D-2}}G_{\mn}$, in this case $D=4$
and $g_{\mn}$ is flat.} and has non vanishing torsion related to the dilaton
through $H=*d\Phi$. This is precisely the condition for a
supersymmetric axionic instanton. Manifest supersymmetry
has been preserved in \aledual~since the relevant
Killing vector is triholomorphic, \ie~it preserves the
hyperk\"ahler structure of \multicenter.
\REF\dix{L.~Dixon, {\it Some World-sheet Properties of Superstring
Compactifications on Orbifolds and Otherwise},
in the Proceedings of the ICTP Summer Workshop in High Energy Physics
and Cosmology, Trieste, Italy 1987.}

\REF\bs{I.~Bakas and K.~Sfetsos, {\it T-duality and world-sheet supersymmetry},
CERN-TH/95-16, hep-th/9502065.}
Buscher's dualities with respect to non triholomorphic isometries yield
string backgrounds
in which supersymmetry is non-locally realized [\bs]. This may force one
to reconsider the relation between spacetime and world-sheet
supersymmetry in superstring compactifications [\dix].

\chapter{Calabi-Yau compactifications and Mirror Symmetry}

\REF\gep{D.~Gepner, {\it Lectures on N=2 String Theory}, in the Proceedings
of the Spring School on Superstrings, ICTP Trieste, Italy 1989.}
Toroidal compactifications of superstring theories bring too many
supersymmetries in $D=4$.
Although supersymmetry is phenomenologically well motivated as the
only known solution of the hierarchy problem, chiral fermions are compatible
only with $N=1$ supersymmetry.
Compactifications of the heterotic or type I superstring which preserve only
one supersymmetry in $D=4$ require six-dimensional backgrounds which admit
only one covariantly constant (Killing) spinors. The relevant
compact manifolds are known as Calabi-Yau spaces (CYS),
which are complex K\"ahler
manifolds of $SU(n)$ holonomy, \ie~vanishing first Chern class.
In the case of interest for superstring phenomenology the number of complex
dimensions is $n=3=6/2$. (Lower dimensional cases are very tightly restricted.
For $n=1$, only the flat torus $T^2$ would do the job. For $n=2$,
manifolds of $SU(2)$ holonomy are hyperk\"ahler and, in the compact case,
they are topologically equivalent to the K\"ummer's third surface $K3$
while, in the non-compact case,
they belong to the class of ALE instantons encountered previously.)
Simply connected manifolds of $SU(3)$ holonomy, ${\cal M}$, are topologically
characterized
by only two independent Hodge numbers: $h_{11}=dimH^{(1,1)}$ (the number of
harmonic (1,1)-forms on ${\cal M}$) and $h_{21}=dimH^{(2,1)}$
(the number of harmonic (2,1)-forms ${\cal M}$).
The other non-vanishing Hodge numbers are $h_{00}=h_{33}=1$,
related to the constant scalar and to the volume top-form,
and $h_{30}=h_{03}=1$, related to
the existence of one (anti)-holomorphic no-where vanishing 3-form.
The Euler characteristic $\chi({\cal M})=2(h_{11} - h_{21})$ turns out to be
twice the net number of generations of chiral multiplets in the {\bf 27}
(${\bf 26}_{(2)}+{\bf 1}_{(-4)}$) of the
low-energy gauge group $E_6\times E_8$ ($SO(26)\times U(1)$) resulting from
the standard embedding of the $SU(3)$ spin connection in the $D=10$ gauge
group.

The massless fluctuations of the metric and of the antisymmetric tensor
correspond to
$h_{11}$ complex parameters of deformations of the (complexified) k\"ahler
class and to
$h_{21}$ complex parameters of deformations of the complex structure. From the
low-energy $D=4$ viewpoint these are complex scalars with only derivative
interactions, known as ``moduli" of the CYS [\dix,\gep].
The same scalars (accompanied by other ones arising from the R-R sector)
would appear in type II string compactification on CYS.
Compatibility with
$N=2$ supersymmetry, then imply that the non-linear $\s$-model
for the moduli fields $\phi$ is based on a ``special K\"ahler manifold"
${\cal S}$
whose k\"ahler potential admits an analytic prepotential ${\cal F}(\phi)$.
Furthermore the $\s$-model manifold has locally a product structure
${\cal S}={\cal S}_{11}\times {\cal S}_{21} \times SU(1,1)/U(1)$,\foot{The last
factor includes the dilaton and the axion, related to $B_{\mn}$ through
$\de_\m b = \epsilon_\m{}^{\n\r\s} \de_\n B_{\r\s}$.}
with $dim_C({\cal S}_{11})=h_{11}$ and $dim_C({\cal S}_{21})=h_{21}$.
The metric on ${\cal S}_{21}$ is neither renormalized by quantum
perturbative corrections in $\ap$ nor by world-sheet non-perturbative effects.
At the classical level, the metric on ${\cal S}_{11}$
is completely determined by the intersection form among three
(1,1)-forms, which is a topological invariant and as such receives no
perturbative corrections in $\ap$. However world-sheet instantons do correct it
non-perturbatively in $\ap$ ($\sim e^{-{R^2\over \ap}}$, where $R^2$ is the
typical size of the CYS). Fortunately ``mirror symmetry" comes
to rescue us.

\REF\gpalr{B.~Greene and M.~Plesser, Nucl. Phys. {\bf B338} (1990) 15;
P.~Aspinwall, A.~L\"utken and G.~Ross, Phys. Lett. {\bf B241} (1990) 373.}
\REF\cdgp{P.~Candelas, X.~De la Ossa, P.~Green, and L.~Parkes,
Nucl. Phys. {\bf B359} (1991) 21;
P.~Candelas, X.~De la Ossa, A.~Font, S.~Katz, and D.~Morrison
Nucl. Phys. {\bf B416} (1991) 21.}
Mirror symmetry emerges from two rather different observations.
The first is the ``experimental" observation that the known CYS
tend to come in pairs with opposite Euler characteristic (\ie~with
$h_{11}$ and $h_{21}$ exchanged). The second is a result from two-dimensional
$N=(2,2)$ superconformal field theory: the projection of the spectrum on
states of integer $U(1)$ charges, necessary to achieve spacetime supersymmetry
[\dix, \gep], allows to revert the $U(1)$
charges in one of the two sectors [\gpalr]. This operation turns out to
interchange deformations of the complex structure with deformations of the
k\"ahler class. One is thus lead to conjecture that any
reasonable Calabi-Yau manifold ${\cal M}$ should
admit a ``mirror image" ${\cal W}$
with $h_{11}({\cal W}) = h_{21}({\cal M})$ and
$h_{21}({\cal W}) = h_{21}({\cal M})$. The appropriate setting to discuss
mirror symmetry is ``toric geometry". For CYS embedded in
toric varieties mirror symmetry should correspond to the exchange of the
``fan" defining the toric variety with its dual\foot{Claudio Procesi,
private communication.}.
Using mirror symmetry one can exactly compute the metric on ${\cal S}_{11}$
for ${\cal M}$
by relating it to the metric on ${\cal S}_{21}$ for the mirror manifold
${\cal W}$ [\cdgp].
The determination of the exact mirror map between the two sets of
parameters relies on the solution of Picard-Fuchs equations for the
periods of the holomorphic three-form along a symplectic basis of three-cycles
on CYS [\cdgp]. Matching the asymptotic behaviours and the
``classical monodromies" one can finally sum up the $\ap$ non-perturbative
contributions arising from the world-sheet instantons, \ie~rational curves
holomorphically embedded in the CYS. The resulting
``quantum-corrected" moduli spaces admit the action of discrete groups of
isometries which are the byproduct of the monodromy group of the Picard-Fuchs
equations and may serve as exact symmetries to constrain the string-loop
corrections to the low-energy effective action. In order to get some
information about string non-perturbative effects the above arguments
are not sufficient. A (discrete) symmetry between strong and weak coupling
would certainly be very helpful.

\chapter{S-duality and H-monopoles}

\REF\emd{C.~Montonen, D.~Olive, Phys. Lett. {\bf 72B} (1977) 117;
P.~Goddard, J.~Nuyts and D.~Olive, Nucl. Phys. {\bf B125} (1977) 1.}
\REF\bps{M.~Prasad and C.~Sommerfield, Phys. Rev. Lett. {\bf 35} (1975) 760;
E.~Bogomolnyi, Sov. Phys. J. Nucl. Phys. {\bf 24} (1976) 449.}
\REF\ggpz{A.~Giveon, L.~Girardello, M.~Porrati and A.~Zaffaroni,
Phys. Lett. {\bf B334} (1994) 331; preprint hep-th/9502057}
The Maxwell equations for the electro-magnetic field in vacuo are manifestly
invariant under the interchange of the electric and magnetic fields.
In fact one can expose a continuous invariance under phase transformations
of the complex vector ${\bf E} + i {\bf B}$. This symmetry is clearly broken by
the presence of electric charges or, equivalently, by the absence of magnetic
charges. In order to restore the symmetry one should introduce magnetic
charge and current
densities. Magnetic monopoles in an abelian theory are singular
object and the motion of electric charges in their presence is inconsistent
unless one imposes the Dirac quantization condition $e  g = 2n\pi$.
In non-abelian theories (such as $SU(2)$) with scalars, however,
't Hooft and Polyakov have
shown that non-singular monopole solutions exist with quantized magnetic
charge. Montonen and Olive [\emd] have conjectured the existence of a symmetry
which exchanges an ``electric" theory with coupling $e$ and a
``magnetic" theory with coupling $g= 2\pi/e$.
This strong-weak coupling duality is substantiated by an observation made
by Manton concerning
the scattering of magnetic monopoles. Goddard, Nyuts and Olive have generalized
the conjecture to larger non-abelian group [\emd], for which one needs the
introduction of dual non-abelian groups,
\eg~for $SU(2)$ the dual group is $SO(3)=SU(2)/Z_2$.
Moreover in the presence of a non-vanishing vacuum angle $\theta$
the effective electric charge is shifted. The
strong-weak coupling duality $g^2 \rightarrow 1/g^2$ is then modified
to an infinite dimensional discrete group of $SL(2,Z)$ transformations
called ``S-duality", which in terms of the
complex coupling constant $\tau = 4\pi i /g^2 + \theta /2\pi$ read
$$
\tau\rightarrow {a \tau + b \over c\tau +d}
\eqn\emduality
$$
with $a,b,c,d \in Z$ such that $ad-bc=1$.
S-duality invariance requires the existence of stable dyon solutions,
\ie~solitons with both electric and magnetic charges, subject to the
Schwinger-Zwanziger quantization condition (see \eg~[\dkl]).
The discovery of dyon solutions by Julia and Zee seems to support the
conjecture. However,
the electrically charged quanta are vector bosons while the
magnetically charged solitons are scalars. Moreover, if proper account is taken
of quantum corrections,
the masses of monopoles and dyons are renormalized and their stability
is jeopardized. Supersymmetry can come to rescue the
situation. Indeed in all N-extended supersymmetric Yang-Mills
(SYM) theories
scalar fields $\phi$ in the adjoint representation appear in the vector
multiplets and magnetic charges appear in the central
extensions $Z$ of the supersymmetry algebra.
States which saturate the Bogomolny -
Prasad - Sommerfield (BPS) bound $M\geq |Z|$, between mass $M$ and central
charge $Z$,
lie in (super)short multiplets and are stable against quantum
(perturbative) corrections due to supersymmetry.
In Yang-Mills theory with global $N=4$ supersymmetry the situation is expected
to persist even non-perturbatively. Explicit instanton calculations and the
determination of the free energy in a box with twisted boundary conditions
seem to support this expectation [\ggpz].

\REF\sw{N.~Seiberg and E.~Witten, Nucl. Phys. {\bf B426} (1994) 19;
Erratum ibid. {\bf B430} (1994) 485; ibid. {\bf B431} (1994) 484.}
However, in theories with $N=2$ supersymmetry Seiberg and Witten [\sw] have
shown that non-perturbative effects (instantons) do
correct the analytic prepotential ${\cal F}(\phi)$
and renormalize the mass formula $M=|q\phi + p\de{\cal F}/\de\phi|$. The
Seiberg-Witten solution for ${\cal F}(\phi)$ shows some very peculiar
features. First of all the non-abelian
$SU(2)$ gauge symmetry is never restored in the quantum moduli space,
\ie~there is no place where $<\phi>=0$.
New massless hypermultiplets (monopoles or dyons) appear
in the spectrum at two special points where ${\cal F}$ is singular.
Moreover introducing a mass perturbation, which explicitly breaks
$N=2$ supersymmetry
to $N=1$, causes the monopole (or dyon) to condense and
provides a model for the realization
of confinement (in the resulting $N=1$ SYM theory) as a dual Meissner effect.
The algebraic solution of Seiberg and Witten, based on the introduction of a
``dynamical elliptic curve", whose periods determine the analytic prepotential,
has been generalized to other non-abelian theories, through the study of
periods of ``hyperelliptic curves", and to $N=2$ Super QCD [\sw].
In $N=2$ theories S-duality exchanges different low-energy descriptions of the
same microscopic theory which is thus not self-dual. On the other end $N=4$,
theories as well as other superconformal theories seem to live in a self-dual
phase.

\REF\sen{A.~Sen, Int. J. Mod. Phys. {\bf A9} (1994) 3707;
A.~Sen and J.~Schwarz, Nucl. Phys. {\bf B411} (1994) 35.}
\REF\ht{C.~Hull and P.~Townsend, Nucl. Phys. {\bf B438} (1995) 109;
ibid. {\bf B451} (1995) 525.}
Since $N=4$ SYM theories coupled to $N=4$ supergravity
are the low-energy limit of heterotic string compactified on a six-torus
it is natural to conjecture
that S-duality should hold or even originate in these theories [\sen].
Schwarz and Sen have shown that at generic points of the moduli space
where the gauge group is abelian, \ie~$U(1)^{28}$,
the equations of motion are indeed invariant under $SL(2,R)$ transformations
$$
\lambda\rightarrow {a \lambda + b \over c\lambda +d} \quad
F^{(a)}_{\mn} \rightarrow
(c \lambda_1 + d) F^{(a)}_{\mn} + c \lambda_2 G^{(a)}_{\mn}
\eqn\sduality
$$
with $\lambda=\lambda_1 +i\lambda_2=\Psi + i e^{-\Phi}$ parametrizing the
dilaton-axion system and $G^{(a)}_{\mn} = (R)^a{}_b\tilde F^{(a)}_{\mn}$ with
$R \in SO(6,22;R)$.
In order for the discrete subgroup $SL(2,Z)$ to be effectively realized in
the theory,
Schwarz and Sen have conjectured the existence of a spectrum of BPS saturated
states invariant under S-duality [\sen]. This spectrum of solitons includes
H-monopoles and dyonic black-holes, which may become massless at special
points of the quantum moduli space [\ht]. Some of the required solitons have
been
explicitly found and their stability has been argued on the basis
of mass non-renormalization
in extended supersymmetric theories. A strong argument in favour of
the existence and stability of the needed BPS saturated states is
still lacking and may require a deeper understanding of S-duality.

\chapter{String-String Dualities and Extended Objects}

\REF\bstt{E.~Bergshoeff, E.~Sezgin, Y.~Tanii and P.~Townsend,
Ann. Phys. {\bf 199} (1990) 340.}
In $D=4$ duality between charges and monopoles is a natural symmetry since
the dual of a two-form (the electromagnetic field strength
$F_{\mn}$) is a two-form (the dual field strength
$\tilde F_{\mn} = \mezzo \epsilon_{\mn}{}^{\rho\s} F_{\rho\s}$).
In diverse dimensions things look quite differently. First of all while
a vector potential $A_\m$ naturally couples to ``world-lines" and thus to
pointlike objects, a $(p+1)$-form potential $B_{\m_1..\m_{p+1}}$ naturally
couples to the ``world-volume" of an extended object with $p$ dimensions,
\ie~a $p$-brane, through the interaction term [\bstt, \dkl]
$$
S_p = \int d^{p+1}\xi \epsilon_{\a_1 \a_2 ... \a_{p+1}}
B_{\m_1 \m_2 ...\m_{p+1}} \de_{\a_1} X^{\m_1}
\de_{\a_2} X^{\m_2}...
\de_{\a_{p+1}} X^{\m_{p+1}}
\eqn\pbrane
$$
The field strength of a $(p+1)$-form is a $(p+2)$-form
and in $D$ dimensions
its dual is a $(D-p-2)$-form which is the field strength
of a $(D-p-3)$-form potential naturally coupled to a $(D-p-4)$-brane.
Electro-magnetic strong-weak coupling duality in $D$ dimensions thus
exchanges $p$-branes and $(D-p-4)$-branes [\dkl].

\REF\wittdyn{E.~Witten, Nucl. Phys. {\bf B443} (1995) 85.}
It has been known for a while that
strings are dual to penta-branes in $D=10$ and to strings in $D=6$.
In the latter case, a candidate
dual pair is formed by the heterotic string compactified on a four-torus
and the
type IIA superstring compactified on a K3 surface, the topologically unique
compact manifold of $SU(2)$ holonomy [\ht, \wittdyn].
The basic argument to support this string-string duality is the coincidence of
the low-energy spectrum at generic points of the 80 dimensional
moduli space $SO(4,20)/SO(4)\times SO(20)/SO(4,20;Z)$. At the points
where the abelian gauge symmetry $U(1)^{24}$ is enhanced through the
HFK mechanism in the heterotic string, one expects the
appearence of new massless vector multiplets in the type IIA string.
These states should be charged with respect to the vector fields of
the R-R sector of the spectrum and cannot arise from
perturbative Kaluza-Klein modes, \ie~momentum or winding states.
They are intrinsically non-perturbative solitons which may correspond to
R-R charged black holes [\ht]. Upon further compactification $D=6$ to $D=4$
both theories become $N=4$ supersymmetric and the heterotic string S-duality
gets mapped to type II string T-duality.

\REF\fhs{S.~Ferrara, Harvey, and A.~Strominger, {\it Second Quantized Mirror
Symmetry} hep-th/9505152;
Kachru and C.~Vafa, Nucl. Phys. {\bf B450} (1995) 69.}
\REF\bcdffrsv{M.~Bill\'o, A.~Ceresole, R.~D'Auria, S.~Ferrara, P.~Fr\'e,
T.~Regge, P.~Soriani and A.~Van Proeyen,
{\it A Search for Non-Perturbative Dualities of
Local N=2 Yang-Mills Theories from Calabi-Yau Threefolds},
hep-th/9506075.}
\REF\agnt{I.~Antoniadis, E.~Gava, K.~Narain and T.~Taylor,
Nucl.Phys. {\bf B413} (1994) 162;
{\it $N=2$ type II - heterotic duality and higher derivative F-terms},
IC/95/177, hep-th/9507115.}
\REF\bcov{M.~Bershadsky, S.~Cecotti, H.~Ooguri and C.~Vafa,
Nucl. Phys. {\bf B405} (1993) 279; Comm. Math. Phys. {\bf 165} (1994) 311.}
\REF\prepmb{M.~Bianchi, in preparation.}
Other candidate pairs of dual string theories are the heterotic string
compactified on $K3\times T^2$ and the type II theories (both A and B)
compactified
on a restricted class of CYS which arise as $K3$ fibrations of the sphere
$S^2$ [\fhs].
If true this duality may shed some light on the locally supersymmetric
extension of the Seiberg-Witten solution of $N=2$ SYM theories.
In the locally supersymmetric case the role of the dynamical Riemann
surface is played by the dynamical CYS of compactification
on the type II side [\bcdffrsv]. Relating the two low-energy effective
lagrangians
through the appropriate mirror map one may in principle exactly solve the
theory at least for the terms with two derivatives on the bosonic fields.
Higher derivative couplings such as $R^2 F^{2g-2}$ have been considered in
[\agnt] and are related to an holomorphic anomaly [\bcov].
On the type II side they are determined by a $g$-loop
computation, on the heterotic side they correspond to
one-loop amplitudes. From an effective lagrangian they arise from
non-perturbative (instanton) effects. Indeed chiral selection rules
imply that interactions of the above type receive contribution
from ALE gravitational instantons with topological charge (Hirzebruch
signature)
$\tau=g$ [\prepmb]. \REF\strom{A.~Strominger, Nucl. Phys. {\bf B451} (1995) 96;
B.~Greene, D.~Morrison and A.~ Strominger, ibid. 109.}
Matching the results for model with low-rank gauge groups
on the heterotic side lends support to
Strominger's interpretation [\strom]
of conifold singularities in Calabi-Yau moduli
spaces as the appearence of massless hypermultiplets of R-R charged black
holes on the type II side [\fhs, \agnt].

\chapter{U-duality: Ramond-Ramond Charged Black-Holes as String Solitons}

\REF\sch{J.~Schwarz, {\it An SL(2,Z) Multiplet of Type IIB Superstrings},
hep-th/9509143;
{\it Superstring Dualities}, hep-th/9509148; {\it The Power of M-theory},
hep-th 9510086.}
In order to understand the origin of the $SL(2,Z)$ S-duality symmetry
in $N=4$ $D=4$ effective field theories for the superstrings, it may prove very
helpful to observe that a similar symmetry is already present in the type IIB
theory in $D=10$. Indeed the two scalars present in the massless spectrum
(one from the NS-NS sector, the other from the R-R sector) may be assembled
into a complex scalar whose self interactions are described by a non-linear
$\s$-model on the coset manifold $SU(1,1)/U(1)$. Though a covariant action
for the type IIB theory with self-dual field strength of the 4-form potential
is still lacking, one may sacrifice manifest supersymmetry and work with
a non self-dual theory, for which an action can be written, and impose
the self-duality constraints only on the equations of motion. Then one can show
that the theory is invariant under $SL(2,R)$ transformations, under which the
four-form is inert. This continuous symmetry is expected to be broken to
a discrete symmetry $SL(2,Z)$ [\sen,\sch,\ht].
After toroidal compactification the two type II theories are equivalent and
in $D=4$ their common massless spectrum precisely fits that of $N=8$
supergravity. This theory contains 70 scalars which parametrize the coset
$E(7,7)/SU(8)$. It is remarkable that the scalars which arise from the
compactification
(``moduli fields") and the dilaton (whose expectation value is the string-loop
expansion parameter) can be transformed into one another. Since the massive
perturbative states of the string break the continuous T-symmetry to a
discrete T-duality, one expects the massive perturbative and
non-perturbative states to break the continuous non-compact $E(7,7;R)$
symmetry to a discrete symmetry group $E(7,7;Z)$ which has been termed
U-duality by Hull and Townsend [\ht].

\REF\sag{A.~Sagnotti, in {\it Non-Perturbative Field Theory}, Cargese 1987,
G. Mack et al. eds. Pergamon-Press, 1988.}
\REF\prad{G.~Pradisi, {\it Open - String Theories} in these Proceedings.}
\REF\mbas{M.~Bianchi and A.~Sagnotti, Phys. Lett. {\bf 247B} (1990) 517.}
The U-duality group includes the product of the T-duality
group $SO(6,6;Z)$ and of the S-duality group $SL(2,Z)$ as a ``maximal" subgroup
(recall that $SO(12)\times SU(2)\subset E_7$). In order for $E(7,7;Z)$
to be a symmetry of the spectrum one has to assume the existence of
solitonic states charged with respect to the 16 R-R vector bosons.
Their masses scale as $M_{RR} \approx 1/g$ [\wittdyn].
There are two pathways for truncating the type II theory to $N=4$.
Either through a target-space orbifold
of the six-torus $T^6$ or through a ``world-sheet"
orbifold [\sag]. The latter leads to a type I supersting which includes
open-string states.
For a complete low-energy description one needs to include all the
solitonic states which may become massless at points corresponding to
the enhanced symmetry points on the heterotic string moduli space [\gpr].
\REF\klct{R.~Kallosh, A.~Linde, T.~Ortin and A.~Peet, Phys. Rev. {\bf D46}
(1992) 5278;
M.~Cvetic and A.~Tseytlin,
{\it General class of BPS saturated dyonic black holes as exact superstring
solutions}, hep-th/9510097.}
{}From the four dimensional point of view, these
$N=4$ supersymmetric solitons are charged extremal black-holes
[\ht, \klct] and the very consistency of the theory forces one to treat them as
elementary particles!
{}From the ten dimensional point of view these states may be pictured as
penta-branes wrapped around internal dimensions. When some of the
homology cycles of the internal compact manifold shrink to zero size,
a conifold transition may take place and the solitonic state
whose mass is proportional to the size of the cycle becomes massless [\strom].
If the soliton is charged, the $U(1)^{28}$ abelian group may be enlarged
to a non-abelian group.

\chapter{Heterotization of the Chan-Paton Group}
\REF\car{J.~Cardy, Nucl. Phys. {\bf b324} (1989) 581.}
In $N=4$ $D=4$ heterotic effective actions, there are points of the moduli
space where the abelian gauge group $U(1)^{28}$ gets enhanced to large
non-abelian groups (at most $SO(44)\times U(1)^6$).
In type II string theories one may expect a similar behavior only if
at the corresponding points in the K3 moduli space some non-perturbative
solitonic state (``R-R charged black holes") become massless.
For type I theories the situation is more intricate.
According to Sagnotti [\sag] type I theories can be considered as
``parameter space orbifolds"
of the type IIB theory. It is not our purpose here to review the construction
of open-string descendants of type IIB superstring models which is the subject
of Pradisi's contribution to these Proceedings [\prad], however it is necessary
to recall some relevant facts.

First of all in order to couple open strings to closed unoriented ones the
standard Polyakov perturbative series must be supplemented with the inclusion
of surfaces
with boundaries and/or crosscaps. These may be considered as orbifolds of
closed oriented surface (at particular values of their moduli) with respect to
anticonformal involutions. The fixed points of the involution, if any, are
the boundaries of the resulting ``open" surface. Conformal field theories on
surfaces of this kind are equivalent to conformal field theories on
double-covering surfaces endowed with a $Z_2$-orbifold projection
of the spectrum under the exchange of left-movers with right-movers.
Roughly speaking, this procedure halves the world-sheet symmetries
as well as their target space byproducts. For instance,
in $D=10$ identifying the type
IIB superstring states under the exchange of left and right movers effectively
reduces the massless spectrum to $N=(1,0)$ supergravity. The truncation
of the closed string spectrum encoded in the torus partition function $T$
is implemented by the introduction of the Klein bottle projection $K$.
These two
contributions make up the ``untwisted sector" of the parameter space orbifold
$$
Z_u = {1\over 2} (T + K)
\eqn\untwisted
$$
The role of the ``twisted sector" is played by the open string spectrum
encoded in the annulus partition function $A$ and its projection, the M\"obius
strip~$M$
$$
Z_t = {1\over 2} (A + M)
\eqn\twisted
$$
In standard orbifold construction the twisted sectors
come out with the multiplicity of the fixed points.
Similarly in the parameter-space
orbifolds the open string states may acquire multiplicities associated to their
ends through the so-called Chan-Paton factors.
Consistency requirements
may be deduced transforming the above amplitudes to the transverse channel,
where Klein bottle $\tilde K$, annulus $\tilde A$ and M\"obius strip
$\tilde M$ are to be identified with closed-string amplitudes between boundary
and/or
crosscap states. Since ``half" of the closed-string states have been projected
out of the spectrum, it would be inconsistent if some of them would couple
to the vacuum. The cancellation between boundary and
crosscap contributions to these tadpoles constrains the
Chan-Paton factors and the signs of the projections [\prad].

In order to explicitly solve these constraints, \ie~in order to linearize
the dyophantine equations for the Chan-Paton factors, it proves very useful
to exploit Cardy's proposal [\car]
of associating a boundary state to each sector
of the spectrum. This amounts to expressing the annulus
amplitude in terms of the fusion rule coefficients. This in turn translates
in a transverse channel annulus amplitude which is a sum of
perfect squares. The transverse channel M\"obius amplitude is then fixed
for consistency in each sector of the spectrum
as an appropriate ``square root" of the product of Klein
bottle and annulus amplitudes. Following this procedure one may construct
a web of consistent models of type I superstring in any
dimension, including supersymmetric models in $D=6$ and $D=4$
as well as non-supersymmetric models in $D=10$ [\mbas].

This is not however the whole story.
Indeed the above construction assumed an almost unique $K$.
Sewing constraints for conformal field theories on closed oriented Riemann
surfaces can be extended to surfaces with boundaries and/or crosscaps.
In particular a crosscap constraint can be deduced [\prad].
In many interesting cases
there are several solutions to this constraint which allow to deduce several
different projections of the closed string spectrum.
The procedure is then reversed, the number of allowed boundary states is
reduced and may be inferred from $\tilde M$.
Many new open-string descendants of (world-sheet) left-right symmetric models
can be constructed systematically [\prad].

\REF\gp{M.~Green, {\it Boundary effects in string theory},
CERN-TH/95-252, hep-th/9510016;
J.~Polchinski, Phys. Rev. {\bf D50} (1994) 6041.}
\REF\pw{J.~Polchinski, {\it Dirichlet Branes and Ramond-Ramond Charges},
NSF-ITP-95-122, hep-th/9510017;
E.~Witten, {\it Bound States of Strings and p-Branes},
IASSNS-HEP-95-83, hep-th/9510135; J.~Polchinski and E.~Witten,
{\it Evidence for Heterotic - Type I duality}, hep-th/9510169.}
Recently these ideas has attracted much attention in the string
literature in connection with the observation that non-perturbative string
corrections may be as large as $exp-{1\over g}$ rather than $exp-{1\over g^2}$
as usual in field theory with coupling constant $g$.
Since a boundary contributes one half a handle to the Euler characteristic
of a surface, the
open-string coupling constant is the square-root of the closed string one.
Thus non-perturbative effects of the above kind are naturally generated by the
introduction of boundaries, \ie~by coupling open strings to the closed string
spectrum [\gp]. The way to achieve this coupling in a natural way
is to observe the existence of solitonic solutions in type IIB theories known
as ``orientifolds" or ``D-branes", \ie~solutions with some of the string
coordinates satisfying Dirichlet rather than Neumann boundary conditions [\pw].
The spectrum of these BPS saturated solitons, whose masses scale as
$M_{RR} \approx 1/g$, is expected to reflect the
$SL(2,Z)$ multiplicity of the type IIB strings constructed by Schwarz [\sch].
These strings, labelled by two integers $(r,s)$,
may be thought as solitons of the fundamental type IIB
string (which corresponds to $r=1$ and $s=0$). To support the resulting formula
for the solitonic string tensions
$$
T_{(r,s)}=T_{(1,0)} \sqrt{(r - s \Psi)^2 + e^{-2\Phi}s^2}
\eqn\tension
$$
Schwarz has conjectured the existence of an
underlying ``M-theory" whose physical excitations are p-branes.
Polchinski has suggested that p-brane solitons carrying R-R charges may
be thought as D-branes, \ie~p-branes with Dirichlet boundary conditions
on the transverse coordinates [\pw]. In particular for $p=1$ one finds the
D-strings which can be interpreted as the $(0,1)$ strings above.
Bound states of these D-strings and N-strings
(\ie~strings with standard Neumann boundary conditions)
can be formed to generate
the other $(r,s)$ dy-strings, which may interact via exchange of open
dy-string states [\pw].

In order to properly couple open dy-string states, one has to solve the
crosscap constraint for the conformal field theory describing D-branes
and then fix the Chan-Paton multiplicities imposing the cancellation of
the tadpoles for the unphysical states. The open dy-string states carry both
Chan-Paton and R-R charges. When some of these states become massless
a new mechanism of symmetry enhancement is thus expected to
take place. It deserves the name of ``heterotization" of the Chan-Paton group.
A better understanding of this mechanism would shed some light on the
strong-weak coupling duality between type I and $SO(32)$ heterotic theory
in $D=10$ proposed in [\pw]. A thorough check
of the conjectured duality could be provided by an explicit determination
of the spectrum of dy-string states. Upon toroidal
compactification to $D=4$, new massless states should somewhere appear
to allow the ``heterosis" of the original
Chan-Paton group $SO(32)$ with the R-R group $U(1)^6$.

\chapter{Outlook}

\REF\is{K.~Intriligator and N.~Seiberg {\it Lectures on Supersymmetric Gauge
Theories and Electro-Magnetic Duality} Trieste Summer School '95,
hep-th/9509066.}
String theory seems to undergo a new period of rapid proliferation of ideas
and the fundamental issue of understanding non-perturbative
effects may not be so far from the reach.
The idea that many different looking
theories can be unified in a common more fundamental description may
require the introduction of extended object of different kinds.
The enormous progress in the study of
supersymmetric gauge theories in $D=4$ after the work of Seiberg and Witten
[\sw, \is] and Strominger's interpretation of the conifold transition [\strom]
point in the direction of suggesting
that any singularity in an otherwise physically consistent theory must be taken
as a hint to some relevant phenomena. In this respect the strange role
that open strings seem to play in closed string theories deserves particular
attention [\gp].

\chapter{Acknowledgements}

It is pleasure to thank the organizers of the Meeting for the kind inviation.
It is fair to say that this expanded version of the talk has grown after
fruitful discussions on this rapidly developing subject with
C. Angelantonj, D. Bellisai, E. Bergshoeff, G. Corb\`o, P. Di Vecchia,
F. Fucito, E. Gava, M. Martellini,
G. Pradisi, G. Rossi, A. Sagnotti, Y. Stanev, A. Tanzini,
M. Testa and G. Travaglini. I am particularly indebted to Giancarlo Rossi
for carefully reading the manuscript.

\endpage
\refout
\end